\documentclass{emulateapj}

\usepackage{amsmath}
\usepackage{amssymb}
\usepackage{epstopdf}

\renewcommand{\vec}{\boldsymbol}

\newcommand{\bt}{\ensuremath{\vec{\theta}}}
\newcommand{\veff}{\ensuremath{V_{\mathrm{eff}}}}

\newcommand{\vveff}{\ensuremath{\vec{V}_{\mathrm{eff}}}}

\newcommand{\vvp}{\ensuremath{\vec{V}_{\mathrm{p}}}}
\newcommand{\vvobs}{\ensuremath{\vec{V}_{\mathrm{obs}}}}
\newcommand{\vvscreen}{\ensuremath{\vec{V}_{\mathrm{s}}}}

\newcommand{\scr}{\ensuremath{s}}

\newcommand{\cucm}{\ensuremath{\textrm{ cm}^{-3}}}

\newcommand{\kpc}{\ensuremath{\textrm{ kpc}}}
\newcommand{\AU}{\ensuremath{\textrm{ AU}}}

\newcommand{\mas}{\ensuremath{\textrm{ mas}}}

\newcommand{\s}{\ensuremath{\textrm{ s}}}

\newcommand{\yr}{\ensuremath{\textrm{ yr}}}
\newcommand{\MHz}{\ensuremath{\textrm{ MHz}}}

\newcommand{\kHz}{\ensuremath{\textrm{ kHz}}}
\newcommand{\mHz}{\ensuremath{\textrm{ mHz}}}
\newcommand{\Msun}{\ensuremath{\, M_{\sun}}}

\newcommand{\kms}{\ensuremath{\textrm{ km s}^{-1}}}
\newcommand{\masyr}{\ensuremath{\textrm{ mas yr}^{-1}}}

\newcommand{\fnu}{\ensuremath{f_\nu}}
\newcommand{\ft}{\ensuremath{f_t}}
\newcommand{\barne}{\ensuremath{\bar{n}_e}}

\begin{document}

\title{Deflection of pulsar signal reveals compact structures in the Galaxy}

\shorttitle{Deflection of Pulsar Signal Reveals Structures}
\shortauthors{Hill et al.}

\author{Alex S.~Hill, Daniel R.~Stinebring, Curtis T.~Asplund, Daniel~E.~Berwick, \\
Wendeline~B.~Everett \& Natalie R.~Hinkel}

\affil{Oberlin College, Department of Physics and Astronomy, Oberlin, OH 44074, USA}

\begin{abstract}
We have detected a strong deflection of radio waves from the pulsar
PSR B0834+06 in scintillation observations. Interference between the undeflected
pulsar image and deflected subimages allows single dish interferometry of the
interstellar medium with sub-milliarcsecond resolution. We infer the presence of
scattering structure(s) similar to those that are thought to cause Extreme
Scattering Events in quasar flux monitoring programs: size $\sim 0.2 \AU$
(an angular size of $0.1 \mas$) with an electron overdensity of $\gtrsim 10^3$
compared to the warm ionized medium. The deflectors are nearly stationary in a
scattering screen that is thin ($\lesssim 5 \%$ of the pulsar-observer distance
in extent), is located $70 \%$ of the way from the Earth to the pulsar, and
has been seen consistently in observations dating back $20$ years. The pulsar
scans the scattering screen at a velocity of $110 \kms$ with a detection radius
of $15 \mas$. Pulsar observations such as these --- particularly with a new
generation of low-frequency radio telescopes with large collecting areas ---
hold promise for improving constraints on the poorly understood physical
characteristics and space density of the deflecting structures.
Such observations may also 
prove useful in correcting deviations the deflectors produce 
 in high-precision timing of millisecond pulsars.
 \end{abstract}


\section{Introduction}
The discovery of Extreme Scattering Events \citep[ESEs;][]{fdj+87} in the late 1980s provided evidence for ultracompact
($\lesssim 1 \AU$), ionized objects in the interstellar medium. Despite
more than 15~years of observational and interpretive work, the physical
conditions, origin, and lifetime of these compact refractors are not well
established \citep{rbc87,f94,rlg97,ww98}. Furthermore, their space density and
possible contribution
to the mass distribution in the Galaxy are  poorly known. The basic difficulty
has been detecting events in either quasar or pulsar monitoring
programs \citep{w01}. In order to substantially affect the flux density
of a radio source, the refractor must be nearly aligned with the source.
The refractors are typically located at distances of kiloparsecs, so a
milliarcsecond alignment is required. Thus, only about a dozen ESEs
have been identified, and few of these are easy to model.

The requirement of stringent alignment is relaxed if the intrinsic source size
is small enough to exhibit interstellar scintillation, one of the
observational consequences of coherent multipath scattering from density
inhomogeneities in the medium. Pulsars are the dominant source in this category.
Interference between various parts of the pulsar image gives rise to faint but
observable periodic fringing patterns in pulsar dynamic spectra (radio flux
density as a function of time and observing frequency). These features have a
clear representation in the secondary spectrum (the squared modulus of the
Fourier transform of the dynamic spectrum) and yield information about the
location of the scattering material and the structure of the image \citep{ra82,
hwg85,cw86, wc87, rlg97}. In particular, periodic fringing in the dynamic
fringing gives rise to parabolic \emph{scintillation arcs} \citep{s01,h03,
crs+04, wms+04} in the secondary spectrum, which imply that the dominant
scattering material is localized in a thin screen along the line of sight.

In this Letter, we present evidence for multiple imaging of the pulsar
PSR~B0834+06 with greater detail than a similar event reported by \citet{rlg97}.
We detected secondary spectrum features caused by at least five separate ray
paths throughout a $26$ day observing run. The angular separation
between the undeflected pulsar image and each of the four deflected subimages
grew linearly at a rate consistent with the velocity of the pulsar. This,
combined with multi-frequency data, imply that the multiple imaging is caused by
one or several deflecting structures similar to those that cause ESEs and that
the structures are stationary in the scattering screen. Such objects, if they
trace out underlying neutral material \citep{ww98}, may contain a significant
fraction of the mass of the Galaxy.

\section{Thin Screen Model}

We use a model of scintillation arcs developed in \citet{s01}, \citet{h03},
\citet{crs+04}, and \citet{ wms+04} and summarized here. The essential feature
of the model is that a scintillation arc is caused by scattering in a thin
screen along the line of sight to the pulsar. Interference between a bright core
of the pulsar image and the scatter-broadened halo gives rise to the basic arc,
and the details of the power distribution along the arc provide information
about the power distribution in the pulsar image. Therefore, our secondary
spectra of PSR~B0834+06 at $327 \MHz$ provide a partial image of the pulsar with
an angular resolution of $\sim 0.3 \mas$ and a field of view of $\sim 40 \mas$ (see below).

\subsection{Image features and interference patterns}

We define a coordinate system in the plane of the sky with the pulsar at the
origin, the $\bt_x$-axis pointing along the effective velocity
vector, and $\bt_y \perp \bt_x$. The pulsar effective velocity \citep{cr98},
\begin{equation} \label{eq:veff}
\vveff = (1 - \scr) \vvp + \scr \vvobs - \vvscreen,
\end{equation}
is the apparent velocity of the image of the pulsar through the thin scattering
screen, where \scr\ is the fractional position of the screen from
the pulsar ($\scr = 0$) to the observer ($\scr = 1$), \vvp\ is the transverse
velocity of the pulsar, \vvobs\ is the transverse velocity of the observer, and
\vvscreen\ is the transverse velocity of the screen. Interference between two
arbitrary points in the image plane, $\bt_1$ and $\bt_2$, causes fringing in
the dynamic spectrum with conjugate time and conjugate frequency of
$\ft = - (\bt_2 - \bt_1) \cdot \vveff/(\scr \lambda)$ and 
$\fnu =D (1-\scr) (\bt_2^2-\bt_1^2)/(2 \scr c)$, 
where $\lambda$ is the observing wavelength. The $\fnu$ coordinate is a measure
of differential time delay between pairs of rays, and $\ft$ represents the
temporal fringe frequency of the interference between the two rays or,
alternatively, the differential Doppler shift between them. Interference between
the ray at the origin and points along the $\theta_x$-axis produces a parabolic
scintillation arc defined by $f_\nu = \eta f_t^2$, where the arc curvature
parameter is
\begin{equation} \label{eq:eta}
\eta = \frac{D \lambda^2 \scr (1 - \scr)}{2 c \veff^2}.
\end{equation}
Interference between the origin
and points with non-zero $\theta_y$ places power inside the parabola
because $\fnu \propto \theta_x^2 + \theta_y^2$.
Interference between a bright spot in the
periphery of the image and the the rest of the image produces an inverted
parabola or {\em arclet} with the same $| \eta |$ and a vertex with an $\ft$
coordinate that is related to the $\theta_x$-coordinate\footnote{The 
minus sign in equation~\ref{eq:thetax} and in the defintion of $f_t$ is present because negative $f_t$ values correspond to deflected rays in front of the moving pulsar;  points behind the moving pulsar (past closest encounter to the deflecting structure) produce a positive $f_t$ feature.}
of the bright spot by 
\begin{equation} \label{eq:thetax}
\theta_{x} = - \left( \frac{\scr \lambda}{\veff}\right) f_t.
\end{equation}
Such features are accentuated when the image is elongated along the velocity
vector or the bright spot lies near the $\theta_x$-axis, or both.

\subsection{Screen Location}
For pulsars with measured proper motions and measured or estimated distances, we
can determine the location of the dominant scattering material from the
curvature of the main scintillation arc using equation (\ref{eq:eta}).
PSR~B0834+06 has an estimated distance \citep{cl03} and a moderately well
determined proper motion \citep{las82}: $D = 0.64 \pm 0.08 \kpc$ and
$\mu = 51 \pm 3 \mas \yr^{-1}$. We have measurements of scintillation arcs in
PSR~B0834+06 dating back to 1981 (Stinebring et al. 2005, in preparation).
Those data and the observations presented here are consistent with a value of
$\eta = 0.47 \pm 0.03~{\rm s}^3$ at $327 \MHz$. This results in a value of
$\scr = 0.29 \pm 0.04$ or a screen that is located a distance
$0.46 \pm 0.08 \kpc$ from the observer. 
These parameters result in a conversion between $f_t$ and angle on the sky
of 50~mHz = 24~mas.  Hence, the angular resolution of the secondary spectrum is
(2 x 24~mas / 180 pixels) = 0.27~mas per pixel;  the useful field of view is about 40~mas.

\section{Observations}

\begin{figure}
\plotone{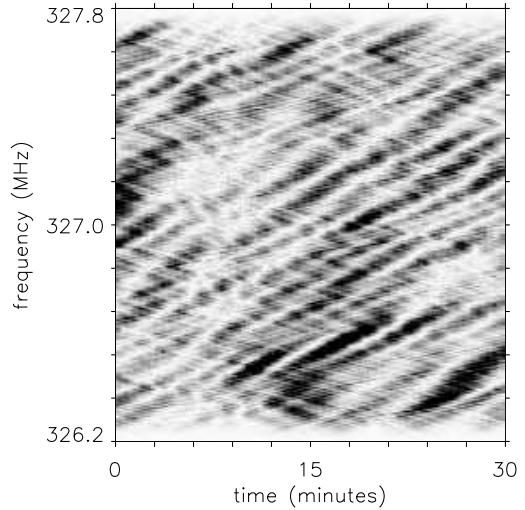}
\caption{The dynamic spectrum of PSR B0834+06 observed on
2003 Dec 31. The flux density as a function of frequency and time is shown
using a grayscale that is linear in power, with dark regions indicating high
power. The crisscross pattern is due to radio waves reaching the observer
from a variety of angles ($\sim$~10~mas away from the pulsar position), as
detailed in the text.}
\label{fig:0834jandyn}
\end{figure}

\begin{figure}
\plotone{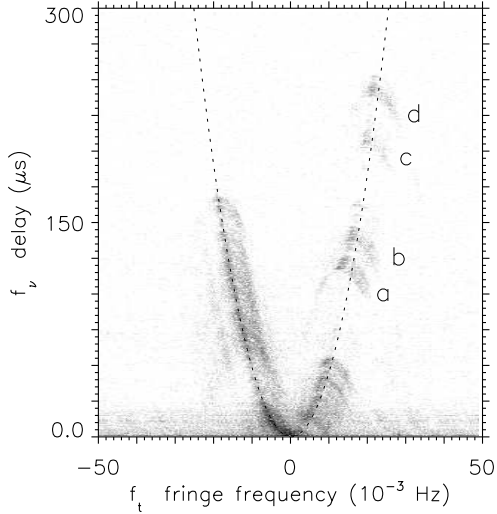}
\caption{The secondary (or delay--fringe-frequency)
spectrum corresponding to the dynamic spectrum in
Figure~\ref{fig:0834jandyn}. The grayscale is logarithmic in power and
represents a total range of about $10^5$ from the maximum power to the noise
level. A primary arc with curvature $\eta = 0.47 \s^3$, fitted to the left side
of the main parabola, is shown with a dashed line. Numerous inverted parabolas,
or \emph{arclets}, are present with vertices near the primary arc.
The four isolated arclets (labeled $a$--$d$) each correspond to a
distinct enhancement in the pulsar image. The horizontal axis can be converted
into a separation angle projected along the direction of pulsar motion using 
$2 \mHz \approx 1 \mas$.}
\label{fig:0834jansec}
\end{figure}

To search for evidence of compact refractors using pulsar scintillation, we
observed the pulsar PSR~B0834+06 at the Arecibo Observatory in 2004 January
using the same technique as \citet{h03}. An example dynamic spectrum is shown in
Figure~\ref{fig:0834jandyn}. We obtained dynamic spectra during 30--60 minute
integrations using the $327 \MHz$ receiver with spectral resolutions of
approximately $1.5 \kHz$ on 11 days over a 26 day period. Typically, we
simultaneously took data with center frequencies of $321$ and $334 \MHz$ using
multiple Wideband Arecibo Pulsar Processor spectrometers. The
secondary spectrum in Figure~\ref{fig:0834jansec} exhibits a scintillation arc
as well as numerous arclets. The four isolated arclets, labeled $a$--$d$,
shifted upward and to the right along the main parabola during the month. This
had never been seen before and is the central observational result of this
paper.

\section{Motion of Arclets}

\begin{figure}
\plotone{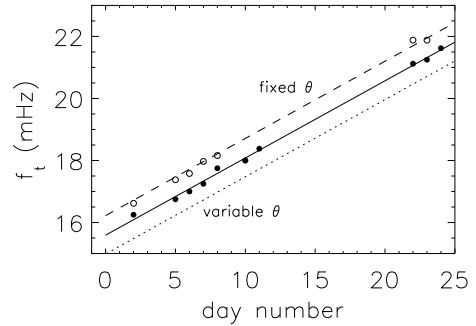}
\caption{The fringe frequency ($f_t$)
coordinate of the vertex of arclet $a$ versus day number (day 0 = 2003 Dec 31).
Data taken at $321 \MHz$ are plotted with filled circles; $334 \MHz$ data are
plotted with open circles, and the uncertainty in position is about the diameter
of the symbols. The solid line represents a two parameter fit to the $321 \MHz$
data. The dashed line has the same slope, but with the offset expected if
$f_t \propto \lambda^{-1}$, or $\theta$ is constant with observing wavelength;
the dotted line has the offset expected if $f_t \propto \lambda$, or scattering
angle $\theta \propto \lambda^2$. These data indicate that  a lens-like
structure or structures in the scattering screen causes the arclets.}
\label{fig:arclet_freq}
\end{figure}

We developed an algorithm to locate the position of these arclets in secondary
spectra by summing along parabolic regions with vertices incremented along the
main parabola. In Figure~\ref{fig:arclet_freq}, we plot the fringe frequency
of arclet $a$ as a function of day number for two closely spaced
frequencies.
Two models of the refracting structures predict different frequency
behavior:  if the rays are refracted by smoothly varying electron density
variations , the angular position of the rays would scale as is
typical for the interstellar medium: $\theta \propto \lambda^2$
\citep[e. g.][]{s68}; then, $f_t \propto \theta \lambda^{-1} \propto \lambda$.
Alternatively,
if the rays giving rise to the arclets pass through a discrete and
dense refracting structure (lens model) \citep{cfl98}, they would arrive from nearly the same
angle $\theta$ at different frequencies; they still deflect as $\lambda^2$ within the structure, but only an unmeasurably slight deviation with wavelength occurs
because of the small size and high density of the refracting structure.
Thus,
$f_t \propto \theta \lambda^{-1} \propto \lambda^{-1}$.
We performed a least-squares linear fit to the lower frequency data. We then
calculated lines with the same slope but vertical offsets  scaling as in the
lens model and in the smooth variation model. The data in
Figure~\ref{fig:arclet_freq} clearly support the lens model, which we assume in
the following discussion.

\begin{figure}
\plotone{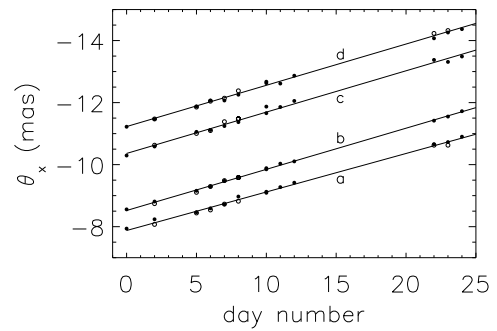}
\caption{The angular position of four arclets over 26
days of observations at two observing frequencies (day 0 = 2003 Dec 31). As in
Figure~\ref{fig:arclet_freq}, filled circles denote $321 \MHz$ data and open
circles indicate $334 \MHz$ data. The uniform, consistent
motion of the arclets is primarily due to the motion of the pulsar, indicating
that arclets are caused by scattering objects essentially stationary in the
screen.}
\label{fig:arclet_pos}
\end{figure}

Using measured values of $f_t$ for the vertices of the arclets, we calculated
their angular position as a function of time, as shown in
Figure~\ref{fig:arclet_pos}. Uncertainties in the angular positions were
typically $0.2 \mas$ as derived from the scatter about the best fit lines.
Arclet $a$ moved at a rate of $47 \pm 2 \masyr$ and arclets $b$--$d$ each
moved at $51 \pm 2 \masyr$ along the pulsar effective velocity vector.
Taking the measured proper motion of the pulsar and accounting for the Earth's
motion at this epoch \citep{kamp,cr98} yields an effective pulsar angular
velocity of $\mu_{\rm eff} = 50 \pm 3 \masyr$. Hence the angular motion of the
pulsar, determined from astrometry, and the rate of increase of the
pulsar--deflector angle, determined from the data reported here, are consistent
within their combined uncertainties. We use this comparison to estimate the
motion of the deflector relative to the screen. We find
$v_{\rm scr} = (1-\scr) D (\mu_\mathrm{arc}-\mu_\mathrm{eff}) \lesssim 9 \kms$,
implying that the physical objects giving rise to the arclets are essentially
stationary in the scattering screen.

\section{Discussion}
The refracting angle of a plasma lens of size $a$ is $\theta_r \sim \alpha
a/D$, where $\alpha \equiv \lambda^2 r_e N_0  D/ \pi a^2$ characterizes the
strength of the refractive effects, $N_0 = \int n_e ds$ is the maximum electron
column density through the lens, and $r_e$ is the classical electron radius
\citep{cfl98}. If we define an average electron density $\barne \equiv N_0 / a$,
then $\barne = (5.4 \cucm) \theta_{\rm r}/\lambda^2$, where the angle is
measured in mas and the wavelength is expressed in meters. For
$\theta_r \approx 15 \mas$ at $327 \MHz$, we find $\barne \approx 100 \cucm$,
which is more than $10^3$ times greater than the average electron density in the
interstellar medium \citep{cl02}. 

The thinness of the arclets is remarkable as well and reinforces the interpretation of refraction by lens-like structures embedded in a thin scattering screen.  
The arclets in Figure~\ref{fig:0834jansec} have a fractional thickness of 
$\Delta \fnu / \fnu \approx$~5\%. Since $\fnu \propto \theta^2$,
this implies 
$a \sim \Delta\theta (1-\scr) D \approx$~0.1~AU if each arclet 
is produced by a separate refractor. In this case the mass in each refractor is
only $M \sim a^3 \barne m_H \approx 10^{-18} \Msun$; here $m_H$ is the mass of
a hydrogen atom and we are assuming a fully ionized object.

Instead of four separate deflecting structures, the four arclets may be caused
by two or even just one structure.
The basic imaging features of a plasma lens become apparent by analyzing a
simple model with a Gaussian column density \citep{cfl98}. This model predicts
caustic surfaces where rays intersect and a defocusing region directly in front
of the lens. In the region between the inner caustic and the outer caustic an
observer would see a bright undeflected image of the source and two dim,
deflected subimages, which would give rise to two arclets as the subimages
interfere with the central image. The angular separation between a pair of
arclets remains constant, as in these observations, for the arclet pairs $ab$
and $cd$, if the refractive strength parameter $\alpha \gg 1$; in this case,
$\Delta \theta \approx 2a / (D - \scr D)$ and
$\barne = \pi \alpha (1 - \scr) \Delta \theta / (2 \lambda^2 r_e) \approx
(1.9 \cucm) \alpha \Delta \theta_\mathrm{mas} / \lambda^2_\mathrm{meter}$.
We can estimate $\alpha$ from the observations because the brightness of the
deflected image relative to the central image is $\approx (1 + \alpha)^{-1}$.
The peak power of the arclets in Figure~\ref{fig:0834jansec} is about $10^{-3}$
of the power at the origin in the secondary spectrum, so $\alpha \gtrsim 10^3$.
Using this model, our observations yield estimates of $a \approx 0.25 \AU$,
$\barne \sim 2000 \cucm$, and $M \approx 4 \times 10^{-16} \Msun$.
It should be noted that neither of these models is in pressure balance in the
warm ionized phase of the interstellar medium; hence, they
would be expected to dissipate quickly unless constrained through some other
mechanism \citep{rbc87, rlg97, ww98}.

\citet{ww98} proposed a model for ESEs in which compact ($\sim 1 \AU$),
long-lived spheroids of neutral hydrogen are surrounded by a photoionized shell.
This results in a double-peaked electron column density profile; each peak may
give rise to two enhancements in the image, resulting in the observed
four-arclet pattern. In this case,
$a \approx \Delta\theta_{ad} (1 - \scr) D/2 \approx 1 \AU$,
where $\Delta\theta_{ad} = 3.3 \mas$ is the separation between arclets $a$ and
$d$. The mass encompassed by the refracting structure is dramatically higher in
this model ($M \sim 10^{-3} M_\odot$) because of the unionized, cold, dense
core. A Galactic halo population of these clouds, sufficient to account for the
poorly constrained ESE rate, would be a major contributor to the total mass of
the Galaxy.

Because interference effects are present even when the deflector does not
intercept the line of sight, pulsar scintillation observations will
improve constraints on the space density of the deflecting structures.
The angular offset of the deflector(s) from the $\theta_x$-axis is an important
parameter in estimating their space density.  Analysis of these data yields
$\theta_y$ offsets in the range $2$ -- $6 \mas$. 
Our sensitivity to this level of offset, along with
the ability to detect deflectors within $10$--$15 \mas$ on either side of the pulsar,
greatly increases the detection probability compared to flux-only observations.

\section{Conclusions}

Scintillation observations provide greater detail on the physical properties of
these refractors than quasar flux monitoring for two principal reasons. First,
the interference between the deflected and undeflected images of the pulsar
allows a single dish telescope to act as an interferometer with excellent
angular resolution. Second, pulsars scan the scattering screen quickly because
of their high transverse velocity. The data presented here demonstrate that
dense, compact refractors similar to those that give rise to ESEs can be
detected in pulsar scintillation observations.
The refracting structures are of interest in their own right, but they also
affect high-precision timing of pulsars \citep{cog93}.  Since this is one of the most
promising methods of detecting a background of primordial gravitational
waves, the ability to detect -- and potentially correct for -- the effects of
refracting structures is of practical importance as well.

We currently do not have enough data containing arclets to decide
between models or to place realistic limits on the space density of refracting
structures. Arclets are not rare, however, particularly at lower frequency
($\lesssim 400 \MHz$) where the effects of scattering and refraction are more
pronounced. We have seen arclets, on occasion, in data from pulsars (PSR)
B0355+54, B0823+26,  B0834+06, B1133+16, B1642--03, B1737+13, and B1919+21.
\citet{rlg97} identified what is probably a poorly-resolved arclet in B0834+06,
and \citet{cw86} and \citet{wc87} report similar features in observations of
B0919+06 and B1237+25. We are analyzing all of these data further in
order to obtain a realistic estimate of the space density of deflectors. It is
clear, however, that single-dish scintillation observations with adequate
sensitivity at low frequency can be used to track deflectors with milliarcsecond
resolution and to estimate their physical properties.

\acknowledgments
The observations were made at the Arecibo Observatory, which is operated
by Cornell University under a cooperative agreement with the National Science
Foundation. The research was supported by National Science Foundation grant
AST 00-98561.

\bibliography{references}

\begin{thebibliography}{22}
\expandafter\ifx\csname natexlab\endcsname\relax\def\natexlab#1{#1}\fi

\bibitem[{Clegg {et~al.}(1998)Clegg, Fey, \& Lazio}]{cfl98}
Clegg, A.~W., Fey, A.~L., \& Lazio, T. J.~W. 1998, \apj, 496, 253

\bibitem[{Cognard {et~al.}(1993)Cognard, Bourgois, Lestrade, Biraud, Aubry,
  Darchy, \& Drouhin}]{cog93}
Cognard, I., Bourgois, G., Lestrade, J.~F., Biraud, F., Aubry, D., Darchy, B.,
  \& Drouhin, J.~P. 1993, \nat, 366, 320

\bibitem[{Cordes \& Lazio(2002)}]{cl02}
Cordes, J.~M. \& Lazio, T. J.~W. 2002, preprint (astro-ph/0207156)

\bibitem[{Cordes \& Lazio(2003)}]{cl03}
---. 2003, preprint (astro-ph/0301598)

\bibitem[{Cordes \& Rickett(1998)}]{cr98}
Cordes, J.~M. \& Rickett, B.~J. 1998, \apj, 507, 846

\bibitem[{Cordes {et~al.}(2004)Cordes, Rickett, Stinebring, \& Coles}]{crs+04}
Cordes, J.~M., Rickett, B.~J., Stinebring, D.~R., \& Coles, W.~A. 2004, \apj,
  submitted (astro-ph/0407072)

\bibitem[{Cordes \& Wolszczan(1986)}]{cw86}
Cordes, J.~M. \& Wolszczan, A. 1986, \apj, 307, L27

\bibitem[{Fiedler {et~al.}(1994)Fiedler, Dennison, Johnston, Waltman, \&
  Simon}]{f94}
Fiedler, R., Dennison, B., Johnston, K.~J., Waltman, E.~B., \& Simon, R.~S.
  1994, \apj, 430, 581

\bibitem[{Fiedler {et~al.}(1987)Fiedler, Dennison, Johnston, \&
  Hewish}]{fdj+87}
Fiedler, R.~L., Dennison, B., Johnston, K.~J., \& Hewish, A. 1987, \nat, 326,
  675

\bibitem[{Hewish {et~al.}(1985)Hewish, Wolszczan, \& Graham}]{hwg85}
Hewish, A., Wolszczan, A., \& Graham, D. 1985, \mnras, 213, 167

\bibitem[{Hill {et~al.}(2003)Hill, Stinebring, Barnor, Berwick, \&
  Webber}]{h03}
Hill, A.~S., Stinebring, D.~R., Barnor, H.~A., Berwick, D.~E., \& Webber, A.~B.
  2003, \apj, 599, 457

\bibitem[{Lyne {et~al.}(1982)Lyne, Anderson, \& Salter}]{las82}
Lyne, A.~G., Anderson, B., \& Salter, M.~J. 1982, \mnras, 201, 503

\bibitem[{Rickett {et~al.}(1997)Rickett, Lyne, \& Gupta}]{rlg97}
Rickett, B.~J., Lyne, A.~G., \& Gupta, Y. 1997, \mnras, 287, 739

\bibitem[{Roberts \& Ables(1982)}]{ra82}
Roberts, J.~A. \& Ables, J.~G. 1982, \mnras, 201, 1119

\bibitem[{Romani {et~al.}(1987)Romani, Blandford, \& Cordes}]{rbc87}
Romani, R., Blandford, R.~W., \& Cordes, J.~M. 1987, \nat, 328, 324

\bibitem[{Scheuer(1968)}]{s68}
Scheuer, P.~A.~G. 1968, \nat, 218, 920

\bibitem[{Stinebring {et~al.}(2001)Stinebring, McLaughlin, Cordes, Becker,
  {Espinoza Goodman}, Kramer, Sheckard, \& Smith}]{s01}
Stinebring, D.~R., McLaughlin, M.~A., Cordes, J.~M., Becker, K.~M., {Espinoza
  Goodman}, J.~E., Kramer, M.~A., Sheckard, J.~L., \& Smith, C.~T. 2001, \apj,
  549, L97

\bibitem[{{van de Kamp}(1967)}]{kamp}
{van de Kamp}, P. 1967, Principles of Astrometry (San Francisco: W. H. Freeman
  and Company)

\bibitem[{Walker(2001)}]{w01}
Walker, M. 2001, Astrophys. \& Space Sci., 278, 149

\bibitem[{Walker \& Wardle(1998)}]{ww98}
Walker, M. \& Wardle, M. 1998, \apj, 498, L125

\bibitem[{Walker {et~al.}(2004)Walker, Melrose, Stinebring, \& Zhang}]{wms+04}
Walker, M.~A., Melrose, D.~B., Stinebring, D.~R., \& Zhang, C.~M. 2004, \mnras,
354, 43

\bibitem[{Wolszczan \& Cordes(1987)}]{wc87}
Wolszczan, A. \& Cordes, J.~M. 1987, \apj, 320, L35

\end{thebibliography}

\end{document}